\documentstyle[prb,aps,twocolumn,floats,epsf]{revtex}
\def\geqap{\,\raise 2pt \hbox{$>\kern-11pt \lower 5pt \hbox{$\sim$}$}\,}
\def\leqap{\,\raise 2pt \hbox{$<\kern-10pt \lower 5pt \hbox{$\sim$}$}\,}
\begin{document}
\draft
\twocolumn[\hsize\textwidth\columnwidth\hsize\csname @twocolumnfalse\endcsname
\title{Interplay of Electron-Phonon Interaction and Electron Correlation 
in High Temperature Superconductivity}
\author{Sumio Ishihara} 
\address{Department of Physics, Tohoku University, Sendai 980-8578, Japan}
\author{Naoto Nagaosa} 
\address{CREST, Department of Applied Physics, University of Tokyo, 
Tokyo 113-8656, Japan \\
and Correlated Electron Research Center, 
Agency of Industrial Science and Technology, Tsukuba, 305-0046, Japan }
\date{\today}
\maketitle
%
\begin{abstract}
We study the electron-phonon interaction in 
the strongly correlated superconducting cuprates. 
Two types of the electron-phonon interactions are 
introduced in the $t-J$ model;  
the diagonal and off-diagonal interactions which 
modify the formation energy of the Zhang-Rice singlet and its transfer integral, respectively.  
The characteristic phonon-momentum $(\vec q)$ and electron-momentum $(\vec k)$ dependence 
resulted from the off-diagonal coupling can explain a variety of experiments. 
The vertex correction for the electron-phonon interaction is 
formulated in the SU(2) slave-boson theory by taking into account 
the collective modes in the superconducting ground states. 
It is shown that the vertex correction enhances the attractive potential for the 
$d$-wave paring mediated by 
phonon with $\vec q=(\pi(1-\delta) , 0)$ around $\delta \cong 0.3$ which corresponds to the half-breathing mode 
of the oxygen motion. 
\end{abstract}
]
%
\narrowtext
%
%
\section{introdcution}
\label{sec:intro}

It is widely accepted that the central issue 
in the high transition-temperature (Tc) superconducting (HTSC) cuprates is 
physics of the doped Mott insulator. 
The Mott-Hubbard (charge-transfer) insulating state 
with the antiferromagnetic long-range order in the parent compounds is well 
understood in strong Coulomb interaction between electrons. 
It is, then, well recognized that 
the electronic models with strong electron correlation provide a good 
starting point to reveal the electronic structure in doped Mott insulator.  
For example, 
one of the successful examples is 
the slave-boson mean-field theory in the two-dimensional $t-J$ model 
and its extensions to the gauge theory which can explain 
a wide range of unusual properties both in the normal and superconducting states; 
the phase diagram in hole concentration $x$ and temperature $T$, 
paring symmetry, pseudogap and so on. \cite{baskaran,kotliar,suzumura,nagaosa,ioffe}

In spite of such successful results in the correlated electronic models, 
one shall have the questions that the electronic model alone is enough 
to explain essential physics in HTSC cuprates. 
Actually, since the early stage of the HTSC researches, 
a variety of experiments exhibit 
the lattice/phonon anomalies and 
the strong electron-lattice coupling which 
stimulate a number of theoretical works. \cite{yonemitsu,ishihara,horsch,nazarenko,normand,boeri}
The finite isotope effects on Tc is one 
of the direct evidence of the  phonon contribution to the paring interaction.  
The oxygen isotope coefficient $\alpha_{O}$ 
in La$_{2-x}$Sr$_x$CuO$_4$ (LSCO) around $x=1/8$ exceeds 0.5 being greater than 
the expected value in the Bardeen-Cooper-Schrieffer (BCS) theory. \cite{crawford}
The isotope shift is also found in  
the magnetic penetration depth at zero temperature $\lambda(T=0)$ (Ref.~\onlinecite{keller}) 
which can not be understood within the standard BCS scenario.
Anomalous phonon dispersion relation is directly 
observed by the inelastic neutron scattering experiments. 
A strong softening and broadening of the highest 
longitudinal optic (LO) phonon  
with the oxygen in-plane Cu-O bond vibrations 
occurs along the $(q_x, 0)$ direction in LSCO. \cite{mcqueeney,pintschovius}
By doping of holes, 
the phonon energy ($\sim$ 85meV) drastically 
reduces about 20$\%$ in the zone boundary, 
and is recovered in overdoped compound. \cite{yamada}

A recently observed kink structure 
in high-resolution angular resolved photoemission spectroscopy  
(ARPES) spectra \cite{bogdanov,kaminski,johnson,lanzara} 
triggers reexamination of the strong 
quasiparticle-phonon interaction in cuprates. \cite{shen_pm,tachiki,hanke,carbotte,rosch}
Here we summarize characteristic nature of this kink structure: 
(a) The kink structure is experimentally confirmed in a variety of the p-type 
cuprates: LSCO, Bi$_2$Sr$_2$CaCu$_2$O$_8$ (Bi2212), 
Bi$_2$Sr$_2$CuO$_6$(Bi2201), (Ca$_{2-x}$Na$_x$)CuO$_2$Cl$_2$ (Na-CCOC),  
Tl$_2$Ba$_2$CuO$_6$ (Tl2201) and so on. 
On the other hand, this structure is weak 
in the n-type Nd$_{2-x}$Ce$_x$CuO$_4$ (NCCO). 
(b) The kink energy $\omega_{kink}$ about 70meV is almost universal. 
The quasi-particle velocity ratio above and below $\omega_{kink}$, i.e. 
$R_v \equiv v(\omega>\omega_{kink})/v(\omega<\omega_{kink})$, is about 2 
in optimal doped p-type cuprates and 
does not show the strong momentum dependence along the fermi surface.  
Below $\omega_{kink}$, the quasi-particle width decreases rapidly. 
This is consistent with the rapid drop of the quasi-particle scattering 
rate deduced from the optical conductivity data. \cite{basov,singley}
(c) The kink structure is seen far above Tc, and  
remarkable changes across Tc are not observed. 
(d) The quasi-particle velocity below the kink $v(\omega<\omega_{kink})$ 
along the nodal direction is almost universal in a wide range of 
HTSC. \cite{zhou} 

This phenomena indicate the strong coupling between 
a bosonic excitation with energy $\omega_{kink}$ and the quasiparticles. 
It is suggested that there is a clear correlation between the 
quasiparticle velocity ratio across the kink $R_v$,  
related to the electron-boson coupling strength, 
and the superconducting gap amplitude. \cite{shen_pm}
A scenario based on the magnetic-resonance mode around 41meV  
as a bosonic excitation \cite{johnson,eschrig,manske}
may be at a disadvantage; this excitation is only confirmed 
below Tc in YBa$_2$Cu$_3$O$_{6+x}$(YBCO), Bi2212 and Tl2201. 
One of the plausible candidates for the  origin of the kink structure 
is the optic phonon. 
The kink energy $\omega_{kink}$ is close to that of a 
LO phonon with the oxygen in-plane vibration 
whose anomalous softening and broadening are observed
as described previously.
This phonon scenario is strongly supported by the recent ARPES experiments 
for the isotope effects on the kink; 
in a Bi2212 sample with the O$^{18}$ isotope, 
the kink energy $\omega_{kink}$ decreases about few meV  
from the standard O$^{16}$ results.  
This result corresponds to the 
simultaneously observed lower shift of the LO phonon 
Raman spectra in the O$^{18}$ sample.\cite{lanzara_iso}  

Here we mention another kink structure in the ARPES 
spectra in Bi2212.\cite{gromko,takahashi,cuk}
This is found near $(\pi, 0)$ and $(0,\pi)$ and, 
its kink energy is about 40meV measured from the Fermi energy. 
It has been claimed that this kink structure almost disappears 
above Tc, 
but recent more extensive study \cite{cuk} found that it
persists even above Tc. 
Therefore the c-axis buckling phonon,
which has the appropriate energy of 40meV, is the most
promising among the various candidates such as 
the magnetic resonance mode and the superconducting gap.
However, we focus below the half-breathing mode
which plays the major role in the nodal region.

In this paper, we study 
the electron-phonon interaction in 
the strongly correlated HTSC cuprates. 
The interactions between electron 
and phonon of the in-plane oxygen vibration mode 
are formulated in the $t-J$ model. 
There are two types of the couplings, i.e. the off-diagonal and diagonal 
interactions which modify the 
inter-site hopping of the Zhang-Rice singlet and its formation energy, 
respectively.
We calculate the effective paring interaction, the quasiparticle 
renormalization factor, the ARPES, tunneling and optical spectra for 
the off-diagonal and diagonal couplings 
where the fermionic degree of freedom is assumed as a quasiparticle with 
the renormalized band energy.  
We show that the characteristic phonon-momentum $(\vec q)$ and 
electron-momentum $(\vec k)$ dependences 
resulted from the off-diagonal coupling are quite consistent with the 
experiments.  
This formulation and numerical results are presented in Sec.~II. 
The vertex correction in the electron-phonon coupling is known to 
play a crucial role in correlated systems. \cite{zeyher,kim}
This is studied, in the slave boson picture, 
by taking into account the fluctuations around the mean-field 
saddle-point solutions, 
that is, the $d$-wave paring order parameter $\Delta$, the fermion 
hopping $\chi$, 
the Lagrangian multipliers $a$ and the bosonic field $h$. 
We formulate this vertex correction in the SU(2) 
slave-boson theory \cite{affleck,wen,lee,brinckmann} 
which respects an exact SU(2) gauge symmetry in the undoped case, and  
is expected to provide a better starting point than the U(1) theory
in the underdoped region. 
The fluctuations are treated as the collective modes in the ordered 
states, \cite{lee2}
and the electron-phonon interactions are corrected by the interaction 
between the collective modes and fermions. 
We show that the vertex correction enhances 
the effective interaction for the d-wave paring 
around $\vec q=(\pi(1-\delta),0)$ with $\delta \cong 0.3$ where 
the remarkable softening and broadening of the LO phonon are 
observed in the inelastic neutron scattering experiments. 
The roles of the vertex corrections are examined in Sec.~III. 
Section IV is devoted to the summary and discussion. 
A part of the present theoretical works were briefly 
discussed in Ref.~\onlinecite{shen_pm}. 

\section{Electron Phonon Interaction in t-J Model}
\subsection{Formulation of the Electron-Phonon Interaction}
We will consider the electron-phonon interaction 
in the $t-J$ model: \cite{zhang}
\begin{eqnarray}
{\cal H}&=&\sum_{i, \sigma} 
\varepsilon_i c_{i \sigma}^\dagger c_{i \sigma}
-\sum_{\langle ij \rangle, \sigma} 
\left ( t_{ij} c_{i \sigma }^\dagger c_{j \sigma}+H.c. \right ) 
\nonumber \\
&+&\sum_{\langle ij \rangle} J_{ij} \vec S_i \cdot \vec S_j , 
\label{eq:tj}
\end{eqnarray}
where $c_{i \sigma}$ is 
the annihilation operator of a hole at site $i$ with spin $\sigma$,  
and is defined in the Hilbert space excluding the double occupancy. 
$\vec S_i$ is the $S=1/2$ spin operator at site $i$. 
The formation energy $\varepsilon_i$ of the Zhang-Rice 
singlet at site $i$ and its 
transfer integral $t_{ij}$ between $i$ and $j$ are represented 
by the energy parameters in the original $p-d$ model as 
\begin{eqnarray}
\varepsilon_i=\frac{2 t_{pd}^2}{\Delta_{pd }(i)} , 
\end{eqnarray}
and 
\begin{eqnarray}
t_{ij}=t_{pd}^2 \frac{1}{2} \left \{ \frac{1}{\Delta_{pd }(i)}+\frac{1}{\Delta_{pd}(j)} \right \} , 
\end{eqnarray}
respectively.  
The charge-transfer energy $\Delta_{pd}(i)$ is defined 
to be the energy difference of the $p$ and $d$ levels at site $i$: 
$\Delta_{pd}(i)=\varepsilon_p-\varepsilon_d(i)$ 
in the hole picture, 
and $t_{pd}$ is the transfer integral between the 
nearest neighboring (NN) $p$ and $d$ orbitals. 
We assume that $U-\Delta_{pd}(i) >> \Delta_{pd}(i)$.  

\begin{figure}
\epsfxsize=0.6\columnwidth
\centerline{\epsffile{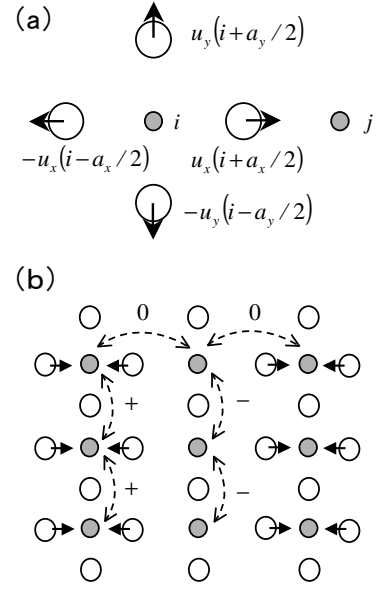}}
\caption{(a) A linear combination of the 
O ion distortion 
$u(i)=u_x(i+a_x/2)+u_y(i+a_y/2)$
$-u_x(i-a_x/2)-u_y(i-a_y/2)$. 
Open and filled curcles indicate the O and Cu ions, respectively. 
(b) Schematic view of the half-breathing mode of the O vibration. 
+, $-$ and 0 indicate signs of the modulation of the 
transfer integral $t_{ij}$ between the NN Cu ions in this displacement. } 
\label{fig:dist}
\end{figure}

Here, we introduce a motion of the oxygen ions 
along the Cu-O bond in the CuO$_2$ plane corresponding 
to the high-frequency optical phonon of our interest. 
The displacement of the negatively charged 
O ion along the Cu-O bond 
changes the electro-static Madelung potential acting on a Cu site and 
shifts the energy of the $d$ level as 
$\varepsilon_d(i)=\varepsilon_{d0}+\delta \varepsilon_d(i)$ with  
\begin{eqnarray}
\delta \varepsilon_d(i)=g u(i) , 
\end{eqnarray}
within the linear electron-lattice coupling. 
$u(i)$ is a linear combination of the O displacements 
around the Cu site $i$ (Fig.~\ref{fig:dist}(a)) defined by 
\begin{eqnarray}
u(i)&=&u_x \left ( i+\frac{a_x}{2} \right )-u_x \left ( i-\frac{a_x}{2} \right )
\nonumber \\
&+&u_y \left ( i+\frac{a_y}{2} \right )-u_y \left ( i-\frac{a_y}{2} \right ) , 
\end{eqnarray}
where $u_l(i \pm \frac{a_l}{2})$ $(l=x,y)$ indicates the 
displacement of the O ion at site 
$i+\frac{a_l}{2}$ along the direction $l$. 
The coupling constant $g$ is positive, 
because the approach of the negatively charged O ion to the 
Cu site lowers the energy level for a hole. 
The O $p$ level is not changed within 
the linear electron-lattice 
coupling, i.e. $\delta \varepsilon_p=0$, because of symmetry. 
The modification of 
the charge-transfer energy is then given by 
$\Delta_{pd}(i)=\Delta_0 - g u(i)$. 
The O ion displacement also changes the 
transfer integral between $p$ and 
$d$ orbitals as 
\begin{eqnarray}
t_{pd}=t_{0} \pm g_t u_l \left ( i \pm \frac{a_l}{2} \right ), 
\end{eqnarray}
for a bond connecting a Cu site $i$ and an O site $i \pm a_l/2$. 
The coupling constant $g_t$ has an opposite sign of $t_0$. 
As a result, 
the modulation of the formation energy of the Zhang-Rice singlet is given by  
\begin{eqnarray}
\varepsilon_i&=& \frac{2 \left \{ t_0+g_t u(i) \right \}^2 }{\Delta_0-gu(i)}
\simeq \frac{2t_0^2}{\Delta_0}+g_{dia} u(i) , 
\end{eqnarray}
with  
\begin{equation}
g_{dia}=\frac{2t_0^2}{\Delta_0} \left(\frac{g}{\Delta_0}+\frac{2g_t}{t_0} \right ) ,  
\end{equation} 
which is termed the diagonal electron-lattice coupling. 
We note that the two terms in the right-hand side have opposite sign with each other. 
In the similar way, 
the transfer integral of the Zhang-Rice singlet is modified as 
\begin{eqnarray}
t_{ij}=\frac{t_0^2}{\Delta_0}+g_{off}\left \{ u(i)+u(j) \right \}, 
\label{eq:tij}
\end{eqnarray}
with 
\begin{eqnarray}
g_{off}=g\frac{t_0^2}{2\Delta_0^2} , 
\end{eqnarray}
termed the off-diagonal 
electron-lattice coupling. 
It is worth noting that 
the modulation of $t_{pd}$ does not change $t_{ij}$. 
This is because, with a shift of a O ion at site $i+a_l/2$ along the 
positive $l$ direction, a decrease in $t_{pd}$ between site $i$ and 
site $i+a_l/2$ is canceled out in the linear order by an 
increase in $t_{pd}$ between $i+a_l$ and $i+a_l/2$. 
The electron-phonon interaction in the $t-J$ model is 
also brought about by the modulation of the 
superexchange interaction $J_{ij}$ 
due to the O displacements. 
This type of electron-phonon coupling was first suggested to explain the 
two-magnon absorption spectra in undoped cuprates
in Ref.~\onlinecite{lorenzana}. 
Here, we consider this electron-phonon coupling in 
the framework of the slave-boson mean-field theory. 
The detailed formulation based on the salve-boson theory is presented in Sec.~III A. 
The constrained operator $c_{i \sigma}$ is  
represented, in this picture, by a product of the spinon (fermion) $f_{i \sigma}$ and holon (boson) $h_{i}$ as 
\begin{equation}
c_{i \sigma}=f_{i \sigma} h_i^\dagger , 
\end{equation}
with the local constraint  
$\sum_{\sigma} f_{i \sigma}^\dagger f_{i \sigma}+h_i^\dagger h_i=1$ 
in order to exclude the doubly occupied states. 
In the mean-field approximation, 
the $J$ term in the $t-J$ model is rewritten by using the 
fermion hopping order-parameter 
$\chi_{ij}(=\sum_{\sigma} f_{i \sigma}^\dagger f_{j \sigma})$ as 
\begin{eqnarray}
J_{ij} \vec S_i \cdot \vec S_j &=& -J_{ij} \chi_{ij}^\dagger \chi_{ij}
\nonumber \\
&\simeq & -\frac{J_{ij} }{2} 
\left ( \langle \chi_{ij}^\dagger \rangle \chi_{ij}
+ \chi_{ij}^\dagger \langle \chi_{ij} \rangle \right ) , 
\end{eqnarray}
which is a similar form to the $t$ term in Eq.~(\ref{eq:tj}). 
The fermion-lattice coupling through the modification of $J$ 
is then reduced to the off-diagonal coupling as 
\begin{eqnarray}
J_{ij}=J_0
+g_J \left \{ u(i)+u(j) \right \} , 
\end{eqnarray}
where the coupling constant is given by 
\begin{eqnarray}
g_J=6 g \frac{t^4}{\Delta_{0}^4} . 
\label{eq:gj}
\end{eqnarray} 

In the standard slave-boson mean-field scheme, where the boson amplitude is taken to be $\sqrt{x}$ in the ground state, 
the diagonal coupling constant $g_{dia}$ and the off-diagonal one 
from the modulation of the transfer integral $g_{off}$ are scaled by the factor $x$, and the 
coupling caused by the superexchange interaction $g_J$ has a factor $1-x$. 
This implies that, in the Mott insulating limit of $x \rightarrow 0$, 
due to suppression of the charge fluctuation, the diagonal coupling becomes irrelevant,
while the off-diagonal type survives because 
the modulation of $J$ affects the spin channel. 
Even at nonzero (but small) $x$ where the diagonal coupling becomes relevant, 
the vertex correction for the electron-phonon interaction, 
induced by the holon fluctuation, suppresses the diagonal coupling 
at the large momentum transfer.\cite{zeyher,kim}
The vertex correction for the off-diagonal coupling is presented 
in Sec.~III in more detail. 

Consequently, 
the electron-phonon interaction Hamiltonian in the $t-J$ model 
is summarized as 
\begin{eqnarray}
{\cal H}_{el-ph}&=&\frac{1}{N} \sum_{\vec k, \vec q, \sigma} 
\sum_{ \mu}
g^\mu(\vec k, \vec q)
\nonumber \\
& \times &
f_{\vec k+\vec q \sigma}^\dagger f_{\vec k \sigma} 
\left ( b_{\vec q}^\mu +b_{-\vec q}^{\mu \dagger} \right ) , 
\label{eq:elph}
\end{eqnarray}
where $b_{\vec q}^\mu$ is the annihilation operator of phonon 
with momentum $\vec q$ and the mode $\mu$.
We take the two independent modes $\mu=x$ and $y$ 
where the O ions vibrate along the $x$ and $y$ directions, respectively.  
The electron-phonon coupling constant is given by a sum of the two-kind couplings  
\begin{equation}
 g^\mu(\vec k, \vec q)=g_{dia}^\mu(\vec k, \vec q)+g_{off}^\mu (\vec k, \vec q), 
\end{equation}
where 
\begin{eqnarray}
g_{dia}^\mu (\vec k, \vec q)=-2i g_{dia}\frac{1}{\sqrt{2MN\omega_{\vec q}^\mu}}
\sin \left (  \frac{q_\mu}{2} \right ) , 
\label{eq:dia}
\end{eqnarray}
and 
\begin{eqnarray}
g_{off}^\mu (\vec k, \vec q)&=&-4i g_{off} 
\frac{1}{\sqrt{2MN \omega_{\vec q}^\mu}}
\sin \left (\frac{q_\mu}{2} \right)
\nonumber \\
&\times & \Bigl \{ 
\cos k_x+\cos k_y 
\nonumber \\
&+&\cos \left (k_x+q_x \right )
+\cos \left (k_y+q_y \right )
\Bigr \} , 
\end{eqnarray}
with the oxygen mass $M$ and the phonon frequency $\omega_{\vec q}^\mu$. 
The off-diagonal coupling constant $g_{off}$ is redefined that 
$g_{off}$ includes the contribution from $g_J$ in Eq.~(\ref{eq:gj}). 
The diagonal-coupling constant $g_{dia}(\vec k, \vec q)$ 
does not depend on the electron momentum $\vec k$ as usual. 
Here, we note the characteristic 
electron-momentum $\vec k$ and phonon-momentum $\vec q$ dependences of the 
off-diagonal coupling $g_{off}^\mu(\vec k, \vec q)$: 
(i)
the fermion degree of freedom does not couple to the two-dimensional 
oxygen breathing mode, i.e. 
$g_{off}^\mu (\vec k, \vec q=(\pi, \pi))=0$. 
This is attributed to the fact that, for the breathing mode,  
$u(i)+u(j)$ in Eq.~(\ref{eq:tij}) becomes zero in all the NN Cu-Cu bonds. 
(ii)  
The coupling constant for the half-breathing mode 
where the O ions shift along $l$ does not depend on 
the $l$ component of 
the electron-momentum, e.g., 
$g_{off}^\mu (\vec k, \vec q=(\pi, 0))$ is independent of $k_x$. 
This is because  
this O ion displacement does not change $t_{ij}$ 
along the direction $l$ but change $t_{ij}$ perpendicular to $l$. 
This characteristics of 
$g_{off}^\mu(\vec k, \vec q)$ are  schematically shown in 
Fig.~\ref{fig:dist} (b). 

\subsection{Physical consequences of the off-diagonal and diagonal 
electron-phonon couplings}
\begin{figure}
\epsfxsize=0.8\columnwidth
\centerline{\epsffile{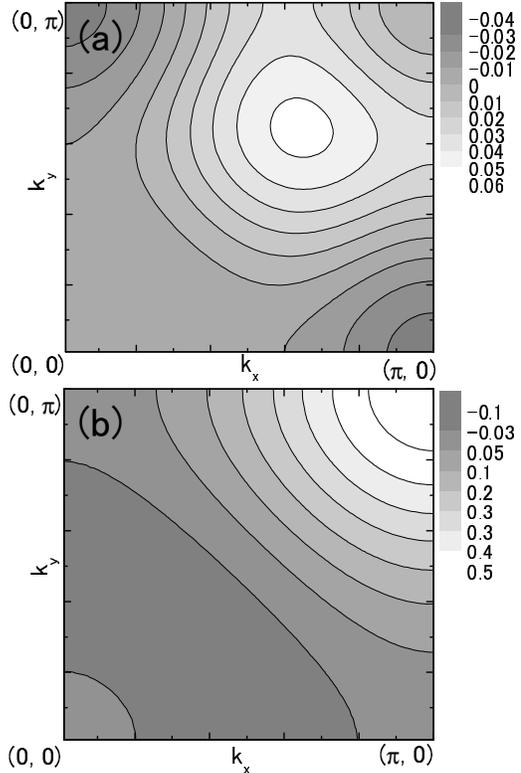}}
\caption{Contour map of the effective paring interaction  
$F(\vec q)$ in the off-diagonal coupling case (a), and in 
the diagonal coupling case (b). 
The energy gap is chosen to be 
$\phi_0 {\widetilde g_{off}}=0.035$eV in (a), 
and 
$\phi_0 {\widetilde g_{dia}}=0.035$eV in (b). 
Numerical data are plotted in a unit of eV. 
The phonon energy is chosen as $\omega_{ph}$=0.07eV. 
} 
\label{fig:fqoff}
\end{figure}
We now turn to the numerical results for various physical 
quantities calculated from 
the electron-phonon interaction Hamiltonian introduced above. 
We pay attention, in particular, to 
the implications of the characteristic momentum dependence of the 
off-diagonal electron-phonon coupling. 
Then, we treat, in this section, an electron as a quasiparticle 
described by the renormalized energy band $\xi_{\vec k}$. 
This treatment corresponds to the mean-field approximation 
in the slave-boson theory where the Lagrangian is given in Eq.~(\ref{eq:lmf}) in Sec.~III A. 
In spite of the detailed form for the quasiparticle energy $\xi_{\vec k}$ 
in the slave-boson theory (see Eq.~(\ref{eq:xi})), 
$\xi_{\vec k}$ is determined by the tight-bind fitting in Bi2212. 
The six tight-binding parameters are adopted by following the results in 
Refs.~\onlinecite{eschrig} and \onlinecite{norman}. 
The fluctuations from the mean-field saddle-points leading 
to the vertex correction for the electron-phonon interaction 
are studied  in the next section. 
The energy parameter values used in this section are given in a unit of eV. 
We assume that the phonon frequency $\omega_{\vec q}^\mu$ is  
independent of the mode $\mu$ and the momentum $\vec q$ as 
$\omega_{\vec q}^\mu=\omega_{ph}$, and choose $\omega_{ph}=$0.07eV. 
The electron-phonon coupling constants are scaled as  
$\widetilde g_{dia(off)}=g_{dia(off)}/\sqrt{2MN\omega_{\vec q}^\mu}$, 
and the damping constant is introduced 
in the calculation of spectra as $\gamma=0.005$eV. 

We first show the 
effective paring interaction for the $d$-wave superconductivity 
due to the electron-phonon interaction. 
The momentum dependence of the effective electron-electron interaction 
is derived by integrating out the phonon degree of freedom 
in the electron-phonon interaction Hamiltonian. 
By using the second-order perturbational processes with 
respect to the electron-phonon coupling, 
the effective Hamiltonian is obtained as 
\begin{eqnarray}
{\cal H}_{eff}&=&-\frac{1}{N} \sum_{\vec k, \vec k', \vec q} \sum_{\sigma, \sigma', \mu}
{g}^{\mu}(\vec k, \vec q) { g}^{\mu}(\vec k', -\vec q)  
\nonumber \\
&\times & 
\frac{\omega_{\vec q}^\mu}
{ \omega_{\vec q}^{\mu 2}-(\xi_{\vec k}-\xi_{\vec k+\vec q})^2 }
f^\dagger_{\vec k +\vec q \sigma}
f_{\vec k \sigma}
f^\dagger_{\vec k' -\vec q \sigma'}
f_{\vec k' \sigma'} .  
\label{eq:heff}
\end{eqnarray}
In the mean-field approximation 
according to the BCS theory, 
we obtain 
\begin{eqnarray}
\langle {\cal H}_{eff} \rangle &=&
-\sum_{\vec q , \vec k, \mu} 
\frac{\omega_{\vec q}^\mu}{ \omega_{\vec q}^{\mu 2}-(\xi_{\vec k}-\xi_{\vec k+\vec q})^2 }
|{ g}^\mu(\vec k , \vec q)|^2 
\phi_{\vec k} \phi_{\vec k+\vec q}
\nonumber \\
& \simeq &
\sum_{\vec q, \mu} \frac{1}{\omega_{\vec q}^\mu} F(\vec q) , 
\label{eq:fq}
\end{eqnarray}
where $|\xi_{\vec k}-\xi_{\vec k+\vec q}| << \omega_{\vec q}^\mu$ is assumed. 
The paring order parameter is introduced as 
$ \phi_{\vec k}=\langle f_{\vec k \uparrow} f_{\vec k \downarrow}\rangle$ 
and is parameterized as $\phi_{\vec k}=\phi_0(\cos k_x-\cos k_y)/2$ 
for the $d$-wave paring. 
$F(\vec q)$ indicates the effective paring interaction between electrons 
contributed from the phonon with momentum $\vec q$. 
In Fig.~\ref{fig:fqoff}(a), 
the contour map of $F(\vec q)$ for the off-diagonal coupling is presented. 
The negative (positive) region corresponds to 
the attractive (repulsive) interaction for the $d$-wave paring. 
It is found that 
the phonon of the momentum $(q_x, 0)$ around $q_x=\pi$ 
strongly contributes to the $d$-wave paring. 
This phonon corresponds 
to the half-breathing mode of the O vibration 
where the anomalous softening of the dispersion relation is observed 
in the neutron experiments.\cite{mcqueeney,pintschovius}  
The vanishing $F(\vec q=(\pi,\pi))$ 
indicates that the off-diagonal electron-lattice 
coupling does not disturb  
the attractive magnetic interaction dominated around $(\pi,\pi)$. 
This originates 
from the characteristic off-diagonal coupling that 
the breathing-mode of the O vibration does not couple to 
the fermion as pointed out above. 
The paring function $F(\vec q)$ 
for the diagonal coupling is presented in Fig.~\ref{fig:fqoff}(b) 
where the attractive interaction appears 
in the region of $q_x+q_y \lesssim \pi$. 
The most remarkable difference from the off-diagonal results is 
seen that the phonon around $(\pi,\pi)$ strongly suppresses
the $d$-wave paring.  
That is, this diagonal coupling competes with the magnetic paring interaction. 

\begin{figure}
\epsfxsize=0.8\columnwidth
\centerline{\epsffile{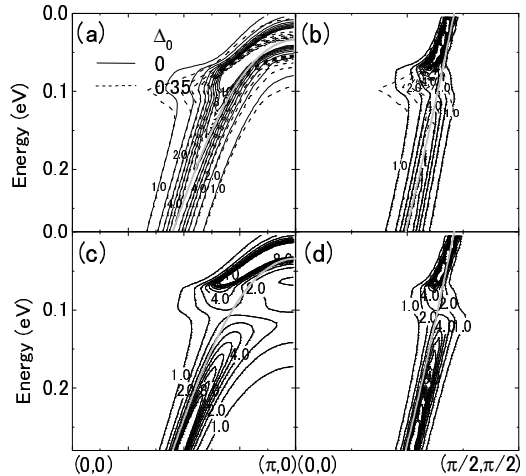}}
\caption{
Contour map of the fermion spectral density $A(\vec k, \omega)$ for 
the off-diagonal case (a) and (b) (${\widetilde g}_{off}$=0.03eV), 
and for the diagonal case (c) and (d) (${\widetilde g}_{dia}$=0.02eV). 
Solid lines indicate the spectra with $\Delta_0=0$ and 
dotted lines in (a) and (b) 
are for the spectra with $\Delta_0=0.035$eV. 
Gray bold lines indicate the bare quasiparticle dispersion $\xi_{\vec k}$. 
The phonon energy is $\omega_{ph}$=0.07eV.
Numerical data are plotted in a unit of eV$^{-1}$.  
} 
\label{fig:spec}
\end{figure}
The ARPES, tunneling and optical spectra are obtained 
in the Nambu-Eliashberg formulae. 
We calculate the self-energy for the 2$\times$2 fermion Green's function:  
\begin{eqnarray}
\Sigma(\vec k, i\omega_n)&=&\sum_{\vec q,  \omega_m,  \mu} \tau_3 G_0(\vec k-\vec q, i\omega_n-i\omega_m) 
\tau_3 
\nonumber \\
&\times &|{ g}^\mu(\vec k, \vec q)|^2 D^{\mu}_{ 0}
(\vec q, i\omega_m) , 
\end{eqnarray}
where we introduce the bare fermion Green's function 
\begin{eqnarray}
G_0(\vec k, i\omega_n)=
\frac{i\omega_n \tau_0+\xi_{\vec k} \tau_3+\Delta_{\vec k} \tau_1}
{(i\omega_n)^2-E_{\vec k}^2} , 
\label{eq:g0}
\end{eqnarray}
with $E_{\vec k}=\sqrt{\xi_{\vec k}^2+\Delta_{\vec k}^{2}}$
and  
the bare phonon Green's function 
\begin{eqnarray}
D^{\mu}_{0}(\vec q, i\omega_m)=\frac{1}{2} \left (\frac{1}{i\omega_m-\omega_{\vec q}^\mu+i \eta}
-\frac{1}{i\omega_m+\omega_{\vec q}^\mu-i \eta} \right ) . 
\end{eqnarray}
$\tau_l$ $(l=1,2,3)$ is the Pauli matrices, $\tau^0$ is a unit matrix, 
and $\Delta_{\vec k}(=\Delta_0 (\cos k_x-\cos k_y)/2)$ 
is the superconducting energy gap. 
The energy and momentum dependences of 
the occupied spectral density corresponding to the ARPES spectra 
are obtained by the (1,1)-component of the Green's function:  
$A(\vec k, \omega)=-\frac{1}{\pi} {\rm Im} G_{11}(\vec k, i\omega_n \rightarrow \omega+i \delta)$.  
The calculated results are presented in Figs.~\ref{fig:spec} (a) and (b) 
for the off-diagonal coupling case, and in (c) and 
(d) for the diagonal case. 
The ARPES spectra show 
a kink structure below which 
both the spectral intensity and the quasi-particle 
lifetime remarkably increase. 
The kink energy in the superconducting state 
is less than $\Delta_0+\omega_{ph}$ because of the $d$-wave gap. 
Although, at a glance, the kink effect in the spectra seems to be remarkable 
in the anti-nodal direction, 
this impression comes from the almost flat dispersion around $(\pi, 0)$ of 
the bare energy band. 
As shown later, the quasiparticle renormalization factor $Z(\vec k, \omega)$ 
dominating 
the velocity change at the kink energy is slightly weaker in the anti-nodal direction 
for the off-diagonal coupling case. 
This is consistent with the recent ARPES experiments where 
the quasiparticle velocity does not show a huge change at $\omega_{kink}$ 
around the anti-nodal direction, unlike the prediction in the magnetic mode 
calculations. \cite{lanzara,eschrig}
\begin{figure}
\epsfxsize=0.9\columnwidth
\centerline{\epsffile{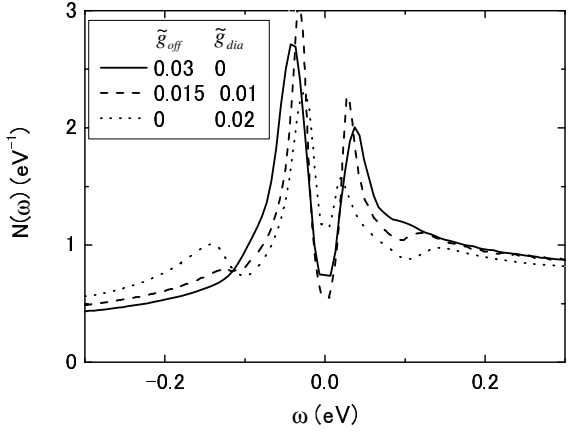}}
\caption{Fermion density of states $N(\omega)$. 
The diagonal and off-diagonal coupling constants $({\widetilde g}_{off}, {\widetilde g}_{dia})$ 
are chosen to be 
(0.03eV, 0) for a solid line, (0.015eV, 0.01eV) for a broken line, and (0, 0.02eV) 
for a dotted line.
Other parameters are  
$\omega_{ph}=0.07$eV and $\Delta_0=0.035$eV.}
\label{fig:dos}
\end{figure}

The tunneling density of states 
\begin{equation}
N(\omega)=-\frac{1}{\pi N} \sum_{\vec k} {\rm Tr \ Im} G (\vec k, i\omega_n \rightarrow \omega+i\delta) , 
\end{equation}
are presented in Fig.~\ref{fig:dos}. 
An asymmetrical shape of the spectra is attributed to the 
almost flat band around the $(\pi,0)$ point below the fermi energy. 
Outside the superconducting gap, 
a dip and hump structure appears, being consistent with experiments. 
\cite{dewilde,renner} 
In the off-diagonal coupling case, 
the weak hump structure is 
located at $\pm(\Delta_0+\omega_{ph})$. 
With increasing the diagonal-coupling parameter, 
this structure becomes pronounced, in particular, 
in the negative side, 
and the position of the structure shifts to the lower energy. 
It is seen simultaneously that for the strong diagonal coupling,  
the gap structure becomes shallow and is gradually collapsed. 
\begin{figure}
\epsfxsize=0.9\columnwidth
\centerline{\epsffile{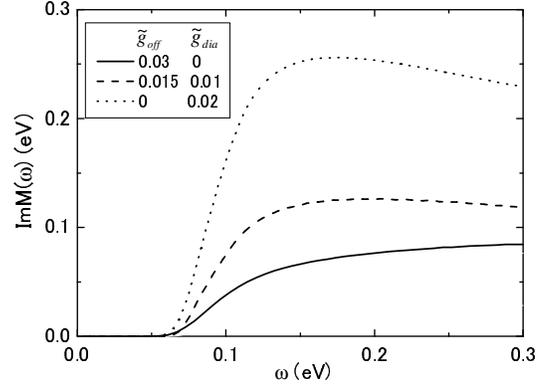}}
\caption{Imaginary part of the 
memory function Im$M(\omega)$. 
The diagonal and off-diagonal coupling constants $({\widetilde g}_{off}, {\widetilde g}_{dia})$ 
are chosen to be 
(0.03eV, 0) for a solid line, (0.015eV, 0.01eV) 
for a broken line, and (0, 0.02eV) 
for a dotted line.
Other parameters are $\omega_{ph}=0.07$eV, $\Delta_0$=0 and $t=0.3eV$.
We use that $\chi_0=0.81ta^2$ calculated for 
the half-filled tight-binding model in the two-dimensional square lattice 
with NN electron hopping. 
}
\label{fig:mem}
\end{figure}
\par
The quasiparticle scattering rate deduced from the optical conductivity 
is calculated in the memory-function formalism. 
The optical conductivity is represented by utilizing the memory function $M_{\alpha \alpha}(\omega)$ 
which corresponds to the inverse of the energy-depend life-time $\tau(\omega)$ as  
\begin{equation}
\sigma_{\alpha \alpha}(\omega)=\frac{i}{4\pi} \frac{\omega_P^2}{\omega+M_{\alpha \alpha}(\omega)} , 
\end{equation}
with the plasma frequency $\omega_P$ and the Cartesian coordinate $\alpha$. 
The memory function $M_{\alpha \alpha}(\omega)$ is defined by \cite{gotze}
\begin{eqnarray}
M_{\alpha \alpha}(\omega)= \frac{1}{ \chi_0 \omega} 
\left \{ \chi_{\alpha \alpha}(\omega)-\chi_{\alpha \alpha}(0) \right \} . 
\label{eq:memo}
\end{eqnarray}
$\chi_{\alpha \alpha}(\omega)$ is the Fourier transform 
of the retarded Green's function of the operator $\vec A_\alpha$ 
defined by the equation of motion for the current operator $j_\alpha$ 
as $ A_\alpha=[ j_\alpha, H]$. 
Explicitly, 
\begin{eqnarray}
A_\alpha&=&-2ta \sum_{\vec k, \vec q, \sigma, \mu} 
\left \{ \sin a (k_\alpha+q_\alpha) -\sin a k_\alpha \right \}  
\nonumber \\
&\times& g^\mu(\vec k, \vec q) f_{\vec k+\vec q \sigma}^\dagger f_{\vec k \sigma} 
\left ( b_{\vec q }^\mu+b_{-\vec q }^{\mu \dagger} \right ) , 
\end{eqnarray}
where we consider the current between the NN sites.
In the calculation of $M_{\alpha \alpha}(\omega)$,
the correlation functions of fermion and phonon operators are
evaluated for the noninteracting Hamiltonian.
$\chi_0$ in Eq.~(\ref{eq:memo}) 
is the static limit of the response function of $j_{\alpha}$. 
The calculated Im$M(\omega)$ are shown in Fig.~\ref{fig:mem}. 
Im$M(\omega)(\equiv $Im$M_{xx}(\omega)=$Im$M_{yy}(\omega))$ is rapidly depressed below $\omega_{ph}$. 
Introducing the diagonal coupling, the scattering rate increases above the threshold energy 
and its reduction below the energy becomes steep. 
This large depression in the quasiparticle scattering rate deduced from the optical spectra 
is commonly observed in a variety of HTSC, e.g. LSCO, YBCO, Tl2201 and NCCO around 
$\omega=400 \sim 800$$\rm cm^{-1}$ (Ref.~\onlinecite{singley}). 
This threshold energy is rather larger than the so-called pseudogap estimated 
in the inelastic neutron scattering, ARPES and tunneling spectra, 
and is close to the kink energy related to $\omega_{ph}$ in the present scenario.  
Absolute values of the calculated Im$M(\omega)$ are compared with the experimental 
quasiparticle scattering rate. 
$\chi_0$ is estimated in the two-dimensional 
tight-binding model with the NN electron hopping. 
Around the threshold energy, the experimental $\tau(\omega)^{-1}$ is about 2000cm$^{-1}$ in 
underdoped Bi2212 and LSCO with $x=0.14$, 1000cm$^{-1}$ in NCCO, and 
500cm$^{-1}$ in overdoped Pb doped Bi2212.\cite{singley}
These values are comparable to the present results of Im$M(\omega)$ in Fig.~\ref{fig:mem}. 
That is, the electron-phonon interaction provides a large portion of the 
scattering rate in the optical region. 

\begin{figure}
\epsfxsize=0.95\columnwidth
\centerline{\epsffile{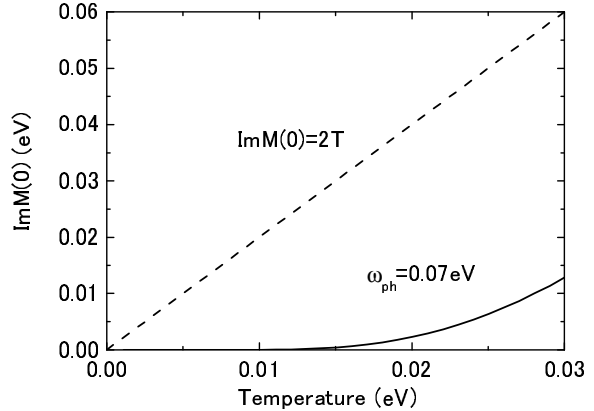}}
\caption{
Temperature dependence of the imaginary part of the memory function 
Im$M(\omega)$ at $\omega=0$ (bold line). 
Broken line indicates a curve of Im$M(0)=2T$. 
The diagonal and off-diagonal electron-phonon 
coupling constants, (${\widetilde g}_{off}$, ${\widetilde g}_{dia}$), 
are chosen to be (0.03eV, 0).
Other parameters are $\omega_{ph}=0.07$eV, $\Delta_0$=0 and $t=0.3$eV.
We use that $\chi_0=0.81ta^2$ calculated for 
the half-filled tight-binding model in the two-dimensional square lattice 
with NN electron hopping. 
} 
\label{fig:resist}
\end{figure}

The memory function Im$M(\omega)$ at $\omega=0$ 
corresponds to the scattering 
rate in the DC reisistivity. 
The temperature dependence of Im$M(\omega=0)$ is presented in Fig.~\ref{fig:resist}. 
It is known that the scattering rate deduced from the experimental resistivity 
data is of the order of 2$T$.\cite{takagi} 
The calculated Im$M(0)$ gradually increases around $\omega_{ph}/5$ 
with increasing temperature, and its absolute value is rather smaller 
than $2T$. 
These results indicate that the phonon contribution is less dominant 
for $\tau(0)^{-1}$ in the resistivity, in contrast to the scattering rate in the optical  
region as mentioned above.  
It is noted that the superconducting gap $\Delta_0$ is put to be zero 
in Figs.~5 and 6. 
When we introduce the finite $\Delta_0$ to describe the pseudo-gap state, 
Im$M(0)$ will be further reduced 
due to the reduced final density of state. 

\begin{figure}
\epsfxsize=0.75\columnwidth
\centerline{\epsffile{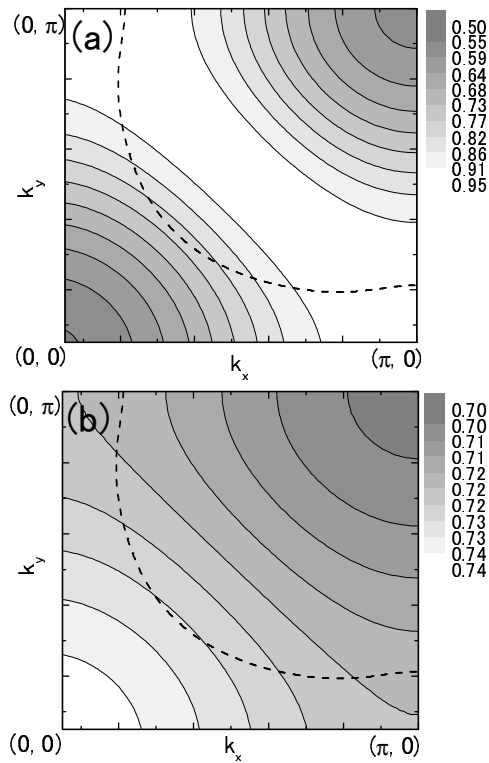}}
\caption{Contour map of the real part of the fermion 
renormalization factor Re$Z(\vec k, \omega=\Delta_0)$ 
in the off-diagonal coupling case (a), and in 
the diagonal coupling case (b). 
Broken lines indicate the fermi surface 
in $\xi_{\vec k}$. 
The diagonal and off-diagonal electron-phonon 
coupling constants, (${\widetilde g}_{off}$, ${\widetilde g}_{dia}$), 
are chosen to be (0.03eV, 0) in (a), and (0, 0.02eV) in (b).
Other parameters are $\omega_{ph}=0.07$eV and $\Delta_0=0.035$eV.}
\label{fig:z}
\end{figure}

The momentum dependences of the fermion renormalization factor 
$Z(\vec k, \omega)=(1-\partial \Sigma(\vec k, \omega)/\partial \omega)^{-1}$ 
at $\omega=\Delta_0$ 
are presented in Figs.~\ref{fig:z}(a) and (b) for the off-diagonal and diagonal couplings, respectively. 
The broken lines indicate the fermi surface for $\xi_{\vec k}$. 
In the off-diagonal coupling case, 
Re$Z(\vec k, \omega=\Delta_0)$ is close to one around the $k_x+k_y=\pi$ line 
in the Brillouin zone, and is strongly 
reduced around $(0,0)$ and $(\pi,\pi)$. 
In contrast, the momentum dependence is much weaker 
in the diagonal case, reflecting that 
the coupling constant $g_{dia}^\mu(\vec k, \vec q)$ is independent of 
the electron-momentum $\vec k$ as seen in Eq.~(\ref{eq:dia}). 
The calculated results for the off-diagonal coupling 
can explain the anisotropic quasiparticle scattering rate 
suggested in the recent ARPES experiments: 
Full width half maximum (FWHM) of the quasiparticle energy distribution curve 
is measured along the fermi surface in Pb-doped Bi2212 compound. \cite{bogdanov2}
The reduction in FWHM on going away from the node is 
observed along both the bonding and antibonding quasi-particle bands. 
This experimental momentum dependence is 
consistent with the off-diagonal results shown in Fig.~\ref{fig:z} (a), 
but is difficult to be explained by 
the magnetic excitation around $\vec q=(\pi, \pi)$ and the diagonal electron-phonon 
coupling. 

\section{vertex correction}
\subsection{Formulation of the Vertex Correction}
We now go beyond the slave-boson mean field treatment in  
the electron-phonon interaction in $t-J$ model. 
This is carried out by taking into account 
the fluctuations 
around the mean fields, 
leading to the vertex correction in the fermion-phonon interaction. 
We formulate this effective electron-phonon coupling in the 
SU(2) slave boson theory \cite{affleck,wen,lee} which imposes 
that the fermion operators 
\begin{eqnarray}
\Psi_{i \uparrow }=
\left ( 
\begin{array}{r}
f_{i \uparrow} \\
f_{i \downarrow}^\dagger
\end{array}
\right ) \ \ , 
\Psi_{i \downarrow}=
\left ( 
\begin{array}{r}
f_{i \downarrow} \\
-f_{i \uparrow}^\dagger
\end{array}
\right )  , 
\end{eqnarray}
are taken to be the local SU(2) doublet. 
This is an exact symmetry in the $t-J$ model at half filling \cite{affleck}
where the two order parameters, i.e.,  
the fermion hopping $\chi_{ij}(=\sum_\sigma f_{i \sigma}^\dagger f_{j \sigma} )$ 
and the fermion paring 
$\Delta_{ij}(= f_{i \uparrow} f_{j \downarrow}-f_{i \downarrow}f_{j \uparrow} )$, 
become equivalent. 
Even at the finite but small doping, 
it is expected that the low energy physics is captured by this picture. 
The generalized SU(2) theory away from the half filling is provided 
by introducing the SU(2) doublet for bosons 
${\hat h}_i=(h_{1i}, h_{2i})^t$. \cite{wen}
Then, the physical electron operator is  
$c_{i \sigma}={\hat h}_i^\dagger \Psi_{i \sigma}$
with constraints 
$\frac{1}{2}\sum_\sigma \Psi_{i \sigma}^\dagger \vec \tau \Psi_{i \sigma}+
{\hat h}_i^\dagger \vec \tau {\hat h}_i=0$. 
Three bosonic fields $a_i^l (l=1 \sim 3)$ 
are required as Lagrange multipliers to keep these constraints. 
It is known that the Lagrangian in the SU(2) theory is rewritten as being similar 
to that in the U(1) theory. This is carried out by representing the two-component bosons 
as a SU(2) rotation, denoted by the rotation matrix $g_i$, 
of the standard slave-boson $h_i$. \cite{brinckmann}
By integrating out the variable $g_i$, 
the Lagrangian for the $t-J$ model in the SU(2) theory associated 
with the electron-phonon 
interaction is obtained as 
\begin{eqnarray}
L=L_{tJ}+L_{el-ph} , 
\label{eq:lag}
\end{eqnarray}
where $L_{tJ}$ is the electronic part:  
\begin{eqnarray}
L_{tJ}&=& \frac{J}{2} \sum_{\langle i j \rangle} {\rm Tr} [U_{ij}^\ast U_{ij}]
+\frac{J}{2}\sum_{\langle ij \rangle, \sigma} \left ( \Psi_{i \sigma}^\dagger U_{ij} \Psi_{j \sigma} +H.c. \right )
\nonumber \\
&+&\frac{1}{2}\sum_{i , \sigma} \Psi_{i \sigma}^\dagger 
\left ( \partial_\tau \tau_0-i a_i^l \tau_l \right ) \Psi_{i \sigma}
\nonumber \\
&+& \sum_i h_i^\dagger \left ( \partial_\tau-ia^3_i +\mu \right ) h_i
\nonumber \\
&-&t\sum_{\langle ij \rangle , \sigma} \left ( 
f_{i \sigma}^\dagger f_{j \sigma}h_i h_j^\dagger +H.c. 
\right ) ,  
\end{eqnarray}
with a 2$\times$2 matrix 
\begin{equation}
U_{ij}=\left ( 
\begin{array}{rr}
-\chi_{ij}^\ast , & \Delta_{ij} \\
\Delta_{ij}^\ast , & \chi_{ij}
\end{array}  
\right ) . 
\label{eq:su2}
\end{equation}
The second term in Eq.~(\ref{eq:lag}) is the 
the fermion-phonon interaction: 
\begin{eqnarray}
L_{el-ph}&=&\frac{1}{\sqrt{N}} 
\sum_{\vec k , \vec q , \mu , \sigma}
g^\mu(\vec k, \vec q) \Psi_{\vec k+\vec q \sigma}^\dagger \tau_3 \Psi_{\vec k \sigma} 
\nonumber \\
&\times&
\frac{1}{2} \left ( b_{-\vec q}^{\mu \dagger} +b_{\vec q}^\mu  \right ) , 
\end{eqnarray}
which is nothing but the Hamiltonian Eq.~(\ref{eq:elph}). 
From now on, we focus on the off-diagonal electron-phonon coupling 
in $g^\mu(\vec k, \vec q)$ of the present interest.  

The fluctuations around the 
mean-field saddle-points in the SU(2) theory 
are examined, in detail in Ref.~\onlinecite{lee},  
as the low-laying collective modes in the superconducting ground state. 
For the fermion hopping and the fermion paring, as well as the 
slave boson and the Lagrange multipliers, 
we introduce small deviations from the saddle-points; 
\begin{eqnarray}
\chi_{ij}&=&\chi_0+\delta \chi_{ij},  
\nonumber \\
\Delta_{ij}&=&\Delta_0(-1) \eta_l +\delta \Delta_{ij} , 
\nonumber \\
h_{i}&=&r_0(1+\delta r_i) , 
\nonumber \\
ia^l_i&=&\delta_{l3}\lambda_0+i \delta a^l_i . 
\end{eqnarray}
$\eta_l=(1,-1)$ for a direction of the bond $l=(x,y)$.  
We adopt the radial gauge where $h_i$ is chosen to be real 
and its phase degree of freedom is absorbed in the Lagrange multipliers. 
Up to the quadratic fluctuations, 
we have a fermionic part of the Lagrangian as 
\begin{eqnarray}
{ L}_{tJ}=L_{MF}+L_1+L_2+L_0 , 
\label{eq:action}
\end{eqnarray}
with the constant term $L_0$. 
$L_{MF}$ is the mean-field part 
\begin{eqnarray}
L_{MF}=-\frac{1}{2N} \sum_{\vec k, \sigma}
\Psi_{\vec k \sigma}^\dagger 
\left ( \partial_\tau \tau_0-\xi_{\vec k} \tau_3 -\Delta_{\vec k} \tau_1 \right )
\Psi_{\vec k \sigma} , 
\label{eq:lmf}
\end{eqnarray}
where we define the mean-field quasiparticle energy  
\begin{eqnarray}
\xi_{\vec k}=-\left ( t_{\vec k}+J \chi_{\vec k}+ \lambda_0 \right ) , 
\label{eq:xi}
\end{eqnarray}
with 
$\chi_{\vec k}=2J\chi_0 (\cos k_x+\cos k_y)$ and 
$\Delta_{\vec k}=2J\Delta_0 (\cos k_x-\cos k_y)$.
The conventional definition of the superconducting gap 
$\Delta$ corresponds to $2J\Delta_0$.  
The bare fermion energy is considered up to the next NN hopping as 
$t_{\vec k}=2r_0^2 \{ t(\cos ak_x+\cos ak_y)+t_1 \cos ak_x \cos ak_y \}$. 
The mean-field solutions, 
$\chi_0$, $\Delta_0$ and $\lambda_0$, are determined 
by the saddle-point equations: 
\begin{eqnarray}
\chi_0&=&-\frac{1}{2N} \sum_{\vec k} 
\frac{\xi_{\vec k} }{E_{\vec k}} \gamma_{\vec k} , 
\\
\Delta_0&=&\frac{1}{N} \sum_{\vec k} \frac{\Delta_0 }{E_{\vec k}} 
J \beta_{\vec k}^2 , 
\\
\lambda_0&=&\frac{1}{N} \sum_{\vec k} 
\frac{2t \xi_{\vec k}}{E_{\vec k}} \gamma_{\vec k} ,  
\end{eqnarray}
with $E_{\vec k}=\sqrt{\xi_{\vec k}^2+\Delta_{\vec k}^2}$, 
$\beta_{\vec k}=\cos k_x-\cos k_y$ and 
$\gamma_{\vec k}=\cos k_x+\cos k_y$.
The calculations in Sec.~III are 
based on $L_{MF}+L_{el-ph}$ where  
$\xi_{\vec k}$ is obtained by the tight-bind fitting 
in spite of Eq.~(\ref{eq:xi}). 
The linear coupling of the fluctuation with the fermionic 
degree of freedom is described by the second term in Eq.~(\ref{eq:action}): 
\begin{eqnarray}
L_1=\frac{1}{\sqrt{2N}} 
\sum_{\vec k ,  \vec q, \sigma} \Psi_{\vec k+\vec q \sigma}^\dagger 
\left ( \vec x_{\vec q} \cdot \vec f_{\vec k, \vec k+\vec q}  \right ) 
\Psi_{\vec k \sigma} . 
\label{eq:l1}
\end{eqnarray}
$(\vec  x_{\vec q})_\nu$ and $(\vec f_{\vec k, \vec k+\vec q})_\nu$ ($\nu=1 \sim 12$) 
are the fluctuations with 12 components and their structure factors, respectively, defined as 
\begin{eqnarray}
\vec x_{\vec q}&=& \Bigl (
\delta \chi_{\vec q x}''  , \  \delta \chi_{\vec q y}''  , \ 
\delta \Delta_{\vec q x}'', \  \delta \Delta_{\vec q y}'', \ 
\delta \Delta_{\vec q x}',  \  \delta \Delta_{\vec q y}',  
\nonumber \\
&\ \ \ & \ \ 
\delta \chi_{\vec q x}', \ \delta \chi_{\vec q y}', \  
\delta a^1_{\vec q},\  \delta a^2_{\vec q}, \  \delta a^3_{\vec q},\  \delta r_{\vec q}
\Bigr ) , 
\end{eqnarray}
and  
\begin{eqnarray}
\vec f_{\vec k, \vec k+\vec q}&=&
\Bigl (
-2J S_x \tau_0, \ -2JS_y \tau_0, \ 2JC_x \tau_1, \ 2JC_y \tau_1,
\nonumber \\
& &
 -2JC_x \tau_2, \ -2JC_y \tau_2,\  
-2JC_x \tau_3, \ -2JC_y \tau_3, 
\nonumber \\ & & 
-i\tau_1, \ -i \tau_2, \ -i\tau_3, \ - \left (t_{\vec k}+t_{\vec k+\vec q} \right )\tau_3 
\Bigr ) , 
\end{eqnarray} 
with abbreviations $S_l=\sin(k_l+q_l/2)$ and $C_l=\cos(k_l+q_l/2)$ for $l=(x,y)$. 
$\delta \chi_{\vec q l}(=\delta \chi_{\vec q l}'+i \delta \chi_{\vec q l}'')$ and 
$\delta \Delta_{\vec q l}(=\delta \Delta_{\vec q l}'+i \delta \Delta_{\vec q l}'')$ are the 
Fourier transforms of 
$\delta \chi_{ij}$ 
and $\delta \Delta_{ij}$, 
respectively, where $l$ indicates a direction of the bond connecting site $i$ and site $j$. 
The third term $ L_2$ in Eq.~(\ref{eq:action}) is the quadratic term of the fluctuations. 
\begin{figure}
\epsfxsize=1.0\columnwidth
\centerline{\epsffile{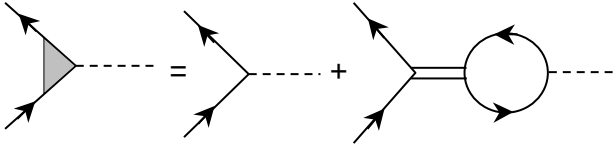}}
\caption{Feynman diagram for the effective electron-phonon interaction vertex.
Bold lines, double lines and broken lines indicate 
the fermioin propagators $G_0(\vec k , i\omega_n)$, 
the bare collective-mode propagators $G_{J0}(\vec q)=M_J(\vec q)^{-1}$, and 
the phonon propagators $D(\vec q, i \omega_m)$, respectively.  
} 
\label{fig:vertex}
\end{figure}
Explicitly,  
\begin{eqnarray}
L_2&=& -2ir_0^2\sum_{\vec k} 
\delta a^3_{\vec k} \delta r_{-\vec k}
\nonumber \\
&-&r_0^2 \sum_{\vec k}
\left (
t \beta_{\vec k} \lambda_1+t_1 \cos ak_x \cos a k_y \lambda_2-\lambda_0 
\right )
|\delta r_{\vec k}|^2
\nonumber \\
&+&J \sum_{\vec k, l=(x,y) } \left ( 
|\delta \chi_{\vec k l}|^2+|\delta \Delta_{\vec k l}|^2
\right ) , 
\label{eq:l2}
\end{eqnarray}
with 
$\lambda_1=- \sum_{\vec k}\frac{\xi_{\vec k} \gamma_{\vec k}}{E_{\vec k}}$ and 
$\lambda_2=- 2\sum_{\vec k} \frac{\xi_{\vec k} \cos ak_x \cos ak_y }{E_{\vec k}}$. 

The effective fermion-phonon coupling is obtained in this scheme.  
The dynamics of the fluctuations $\vec x_{\vec q}$ is governed through the 
interaction between the fluctuations and fermion. 
By integrating out the fermionic degree of freedom in the action, 
we obtain the RPA-type summation 
of the bare Green's function for the fluctuations 
$G_{J0}(\vec q)=F.T. \langle \vec x_{\vec q}(\tau) \vec x_{-\vec q}^\dagger (0) \rangle_0 $, 
where $\langle \cdots \rangle_0$ indicates an expectation in $L_2$, 
and the fermion polarization function 
$M_F(\vec q, i\omega_n)$ defined by 
\begin{eqnarray}
M_F(\vec q, i\omega_n)&=&\sum_{\vec k, \omega_m} \vec f_{\vec k+\vec q, \vec k} 
G_0(\vec k, i\omega_m) 
\nonumber \\
&\times&
G_0(\vec k+\vec q, i\omega_m+i\omega_n)
\vec f_{\vec k+\vec q, \vec k} . 
\end{eqnarray}
$G_0(\vec k, i\omega_m)$ is the bare fermion Green's function 
calculated by $L_{MF}$ (Eq.~(\ref{eq:lmf})). 
The effective fermion-phonon vertex is now given by the bare interaction 
plus the correction due to the collective modes of the fluctuations 
(Fig.~\ref{fig:vertex}). 
\begin{figure}
\epsfxsize=0.85\columnwidth
\centerline{\epsffile{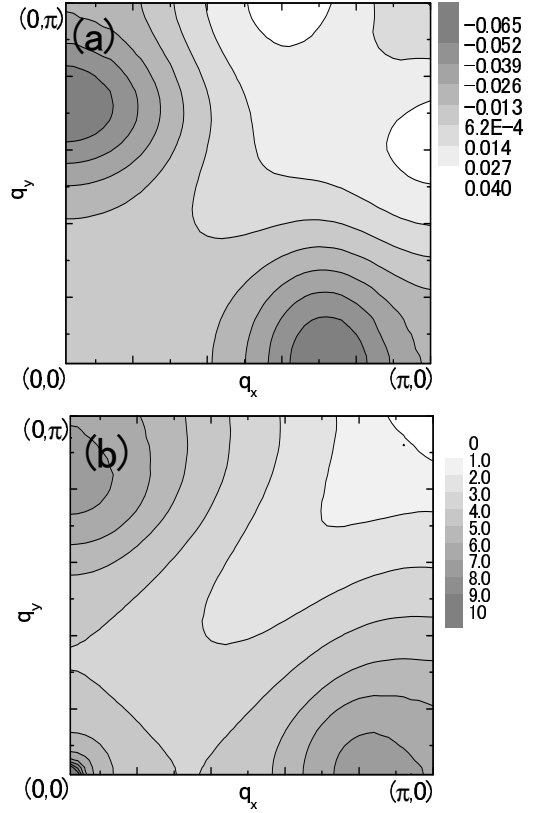}}
\caption{Contour map of (a) the effective paring interaction 
$F(\vec q)$ in the off-diagonal coupling case with the vertex correction, 
and (b) the vertex function $\Gamma(\vec q, 0)$.
Parameters are chosen to be 
$t/J=3$, $t_1/J=-1.5$, ${\widetilde g_{off}}/J=0.1$, $\omega_{ph}/J=0.5$ and $x=0.1$. 
Numerical data are plotted in a unit of $J$. } 
\label{fig:fqvtx}
\end{figure}
Explicitly, 
the coupling constant is 
\begin{eqnarray}
{\widehat g}^\mu(\vec k, \vec q; i\omega_n)&=&
\vec \eta_{\vec q}^\mu \cdot { \Lambda}(\vec q, i\omega_n) \cdot \vec f_{\vec k, \vec k+\vec q} \ . 
\label{eq:geff}
\end{eqnarray}
$\vec \eta_{\vec q}^\mu$ is a vector with 12 components 
defined as 
\begin{eqnarray}
\left (\vec  \eta_{\vec q}^\mu \right )_\nu&=&4i g_{off} J^{-1} \frac{1}{\sqrt{2NM\omega_{\vec q}^\mu}} 
\sin \left ( \frac{q_\mu}{2}\right )
\nonumber \\
& \times & 
\left \{\delta_{\nu 7} \cos \left (\frac{q_x}{2} \right ) 
       +\delta_{\nu 8} \cos \left (\frac{q_y}{2} \right ) \right \} . 
\label{eq:eta}
\end{eqnarray}
The finite components at $l=7$ and 8 in $\eta_{\vec q}^\mu$ are attributed to 
the fact that 
the fermion-phonon vertex is proportional to $\tau_3$. 
${ \Lambda}(\vec q, i\omega_n)$ is the energy-dependent vertex function 
of a 12 $\times$ 12 matrix 
\begin{eqnarray}
\Lambda(\vec q, i\omega_n)=\frac{M_J(\vec q)}{M_J(\vec q)
+M_F(\vec q, i\omega_n)} , 
\end{eqnarray}
represented by the inverse of the bare fluctuation Green's function 
$M_{J}(\vec q)=G_{J 0}(\vec q)^{-1}$ and 
the renormalized fluctuation Green's function 
\begin{equation}
{ G}_{J}(\vec q, i \omega_n)=\frac{1}{M_J(\vec q)+M_{F}(\vec q, i \omega_n)} , 
\end{equation}
which describes dynamics of the collective modes in the 
superconducting ground state.  
\subsection{numerical results with vertex correction}
\begin{figure}
\epsfxsize=0.95\columnwidth
\centerline{\epsffile{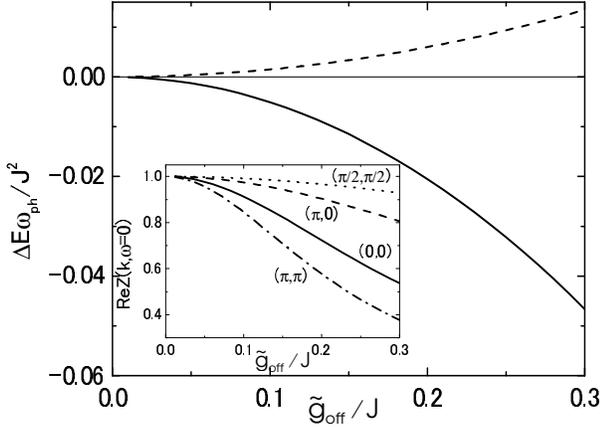}}
\caption{Superconducting condensation energy $\Delta E$.
Solid and broken lines indicate $\Delta E \omega_{ph}/J^2$ 
with and without the vertex correction, respectively. 
The inset shows the real part of the fermion renormalization factor Re$Z(\vec k, 0)$. 
Parameters are chosen to be 
$t/J=3$, $t_1/J=-1.5$, $\omega_{ph}/J=0.5$ and $x=0.1$. 
} 
\label{fig:cond}
\end{figure}
We show, in this section, the numerical results obtained 
by taking into account the vertex correction 
for the electron-phonon interaction in Eq.~(\ref{eq:geff}).
As shown below, the effects of the vertex correction depends strongly on the 
physical quantities. 
Some of them are seriously modified, while others are insensitive to the vertex correction. 
The parameter values are given 
in a unit of $J$, and we use $t/J=3$, $t_1/J=-1.5$, 
${\widetilde g}_{off}/J=0.1$, $x=0.1$ and $\omega_{ph}/J=0.5$. 
The holon is assumed to be condensed and its amplitude $r_0$ 
is taken to be $\sqrt{x}$. 
The obtained mean-filed solutions 
are $\Delta_0$=0.278 and $\chi=0.370$. 

The vertex correction is a crucial for 
the calculation of the paring function and the condensation energy. 
The momentum dependence of the effective paring interaction 
$F(\vec q)$ is obtained by Eq.~(\ref{eq:fq}) where 
$g^\mu(\vec k, \vec q)$ is replaced by the effective coupling constant 
${\widehat g}^\mu(\vec k, \vec q; i \omega_n=0)$ defined in Eq.~(\ref{eq:geff}). 
The numerical results are presented in Fig.~\ref{fig:fqvtx} (a). 
The vertex correction does not only enhance magnitude 
of the attractive interaction, as will be discussed in more detail below, 
but also shifts the minimum region from $(\pi,0)$ to $(\pi(1-\delta), 0)$ 
with $\delta \cong 0.3$. 
This is exactly the momentum region where the minimum of the 
anomalous dispersion is observed in the inelastic neutron scattering. 
\cite{mcqueeney,pintschovius}
We plot, in Fig.~\ref{fig:fqvtx} (b) the contour map of 
the absolute value of the vertex function 
$\Gamma(\vec q, \omega=0)$ defined by 
\begin{eqnarray}
\Gamma(\vec q, i \omega_n)&=&\sum_\nu \biggl \{
   \Lambda(\vec q, i \omega_n )_{7 \nu }   \cos \left (\frac{q_x}{2} \right )
\nonumber \\
&\ &\ \ \ \ +\Lambda(\vec q, i \omega_n )_{8 \nu }   \cos \left (\frac{q_y}{2} \right )
\biggr \} . 
\end{eqnarray} 
We resolve this vertex function into the 
contribution from the collective modes by 
following the classification in Ref.~\onlinecite{lee2}.  
The strong intensity around $\vec q=(0,0)$ in $\Gamma(\vec q, 0)$ is attributed to the Goldstone mode, 
i.e. the uniform collective phase mode of the superconducting order parameter. 
This eigen vector is denoted as 
\begin{equation}
\delta \alpha_{U \vec q} =\delta \Delta_{\vec q x}''-\delta \Delta_{\vec q y}'' ,  
\end{equation}
where $\alpha$ indicates the overall phase of $\chi_{ij}$ and $\Delta_{ij}$ 
around the staggered-flux order parameters. 
This contribution in the effective fermion-phonon interaction 
is, however, irrelevant, because of the factor $\sin(q_\mu/2)$ in Eq.~(\ref{eq:eta}) and also because the long range Coulomb interaction lifts up the
frequency to plasma energy in real system. 
A large contribution around $(\pi, 0)$ is mainly
attributed to the two collective modes, i.e. the amplitude mode  
\begin{equation}
\delta A_{\vec q l}=\Delta_0 \delta \Delta_{\vec q l}'+\chi_0 \delta \chi_{\vec q l}', 
\end{equation}
for $l=(x,y)$ 
representing the amplitude fluctuation for the order parameter  
$|\chi_{\vec q l}|^2+|\Delta_{\vec q l}|^2$, 
and 
the transverse and longitudinal $\phi$ gauge modes 
\begin{eqnarray}
\phi_{T \vec q}&=&\chi_0 \delta \Delta_{\vec q y}'-\Delta_0 \delta \chi_{\vec q y}' , 
 \\
\phi_{L \vec q}&=&\chi_0 \delta \Delta_{\vec q x}'+\Delta_0 \delta 
\chi_{\vec q x}' .
\end{eqnarray}
The amplitude fluctuation $\delta A_{\vec q l}$ has a large intensity 
around $(\pi,0)$ and $(0,\pi)$, and increases the effective paring 
interaction $F(\vec q)$ around these momenta.  
This may be a remnant of the instability toward the columnar-dimer state 
which is well known in the insulating case. \cite{marston,read}
The $\phi$ modes are the amplitude modes with the out-of-phase fluctuation 
for the fermion hopping $\chi_{ij}$ and paring $\Delta_{ij}$. 
The longitudinal $\phi$ gauge mode $\phi_{L \vec q}$ 
shows a large intensity in the low energy region around $\vec q=(\pi/2, 0)$, 
in addition to the peak structure 
around $\vec q=(\pi/2, \pi)$ suggested in Ref.~\onlinecite{lee2}. 
This mode shifts the minimum point of $F(\vec q)$ from 
$(\pi,0)$ without vertex correction to $(\pi(1-\delta), 0)$ with $\delta \cong 0.3$ 
in the low energy region. 
\begin{figure}
\epsfxsize=0.9\columnwidth
\centerline{\epsffile{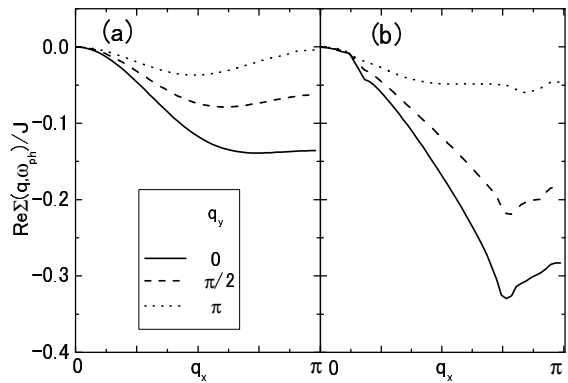}}
\caption{Real part of the phonon self-energy 
without the vertex correction (a), and with 
the vertex correction (b). 
Parameters are chosen to be 
$t/J=3$, $t_1/J=-1.5$, ${\widetilde g}_{off}/J=0.1$, $\omega_{ph}/J=0.5$ and $x=0.1$. } 
\label{fig:phself}
\end{figure}
\par
The superconducting condensation energy $\Delta E$ is given 
by the expectation value of the effective BCS Hamiltonian, i.e. 
Eq.~(\ref{eq:fq}) where 
the vertex correction is included in the coupling constant.  
The numerical results of $\Delta E$ as function of the bare 
off-diagonal coupling constant $g_{off}$ is presented in Fig.~\ref{fig:cond} 
associated with the renormalization factor Re$Z(\vec k, \omega=0)$. 
As shown in the broken line in Fig.~\ref{fig:cond}, 
the condensation energy is positive in the case without vertex correction.  
Thus, the $d$-wave superconductivity due to the present LO phonon 
is stabilized by introducing the vertex correction. 
By considering the velocity ratio $R_v$ across the kink energy, 
a value of $\widetilde g_{off}/J$ is expected to be 0.2 where Re$Z(\vec q, 0)$ is 
about $0.6 \sim 0.8$ at $\vec q=(0,0)$ and $(\pi, \pi)$, 
and $\Delta E \omega_{ph}/J^2=-0.02$ is converted into about 40K, 
taking $J=100$meV. 
The actual condensation energy is of the order of 7K for optimally doped
YBCO \cite{Loram}, which is smaller than this value.
One can consider various reasons for this discrepancy.
(i) the estimation of the condensation energy has an 
uncertainty due to the subtle subtraction of the phonon contribution,
(ii) there are other mechanisms to reduce the pairing and hence the 
condensation energy which are not considered here,
and (iii) the determination of Re$Z(\vec k, 0)$, i.e. 
the electron-phonon coupling constant $\lambda$ from the experimental velocity 
ratio may be overestimated. 
\begin{figure}
\epsfxsize=0.85\columnwidth
\centerline{\epsffile{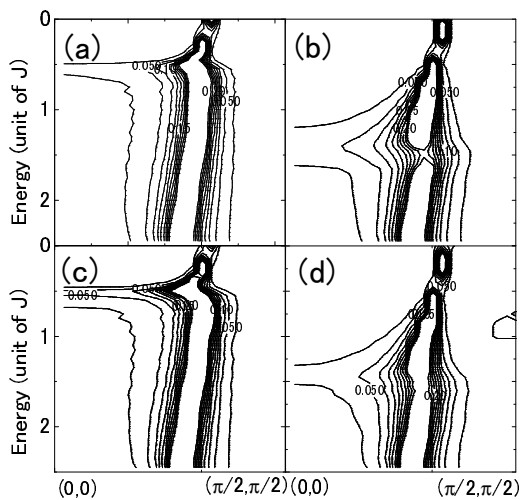}}
\caption{
Contour map of the fermion spectral density $A(\vec k, \omega)$ along the nodal direction 
without the vertex correction (a) and (b), and with the correction (c) and (d). 
(a) and (c) are for the spectra with $\Delta_0=0$, and (b) and (d) are for the 
spectra with $\Delta/J=2\Delta_0/J=0.57$. 
Parameters are chosen to be 
$t/J=3$, $t_1/J=-1.5$, ${\widetilde g}_{off}/J=0.1$, $\omega_{ph}/J=0.5$ 
and $x=0.1$. 
The dispersion relation of the quasi-particle energy $\xi_{\vec k}$ 
is chosen to be the same form with that in Fig.~\protect\ref{fig:spec}. 
Numerical data are plotted in a unit of $J^{-1}$. 
} 
\label{fig:vtxarpes}
\end{figure}

The momentum dependence of the phonon self-energy is presented 
in Fig.~\ref{fig:phself}. 
The phonon self-energy $\Sigma_{ph}(\vec q, \omega=\omega_{ph})$ 
is calculated from the fermion bubble of the bare fermion Green's 
function $G_0(\vec k, \omega)$ associated with the vertex correction. 
We consider the phonon mode where the oxygen ions vibrate along the $x$ direction. 
As shown in the Fig.~\ref{fig:phself}(a),  
the softening of the phonon dispersion is remarkably seen along the $(q_x,0)$ direction 
with $q_y \lesssim \pi/2$. 
This is the region where the softening of the oxygen vibrational LO phonon mode 
is reported in the inelastic neutron scattering experiments. 
The phonon softening is gradually reduced with increasing $q_y$, 
and along the $(q_x, \pi)$ direction, the phonon frequency is almost unrenormalized. 
By introducing the vertex correction (Fig.~\ref{fig:phself}(b)), 
these tendency become significant.
Although the qualitative feature of the calculated phonon softening is consistent with the experiments, 
absolute value of the frequency shift is much 
larger than the observation, in particular with the vertex correction. 
This may indicate that the adopted electron-phonon coupling constant is 
overestimated, as mentioned previously, 
and a further refinement for the vertex correction is required 
for more quantitative discussion. 

On the other hand, the electronic spectra are not sensitive to 
the vertex correction as shown below. 
Figures~\ref{fig:vtxarpes} show the ARPES spectra along the nodal direction 
with and without the vertex correction. 
The energy dependence of the vertex function is taken into account in the calculation of the spectra. 
Comparing Figs.~\ref{fig:vtxarpes} (a), (b) and (c), (d), 
it is concluded that the vertex correction does not play significant roles in the 
ARPES spectra. This is because the electronic self-energy is given by the contributions 
from the wide range of the phonon momentum $\vec q$, and the vertex correction 
is large only in a rather limited $\vec q$ region as shown in Fig.~\ref{fig:fqvtx}(b). 
This fact is also consistent with the shift of the kink position in the superconducting 
state; the kink energy shifts from $\omega_{ph}$ to $\omega_{ph}+2\alpha \Delta$ with $\alpha \cong 0.7$. 
However, this does not reproduce the experimental observation 
that the kink position does not change across the superconducting transition. 

\begin{figure}
\epsfxsize=0.85\columnwidth
\centerline{\epsffile{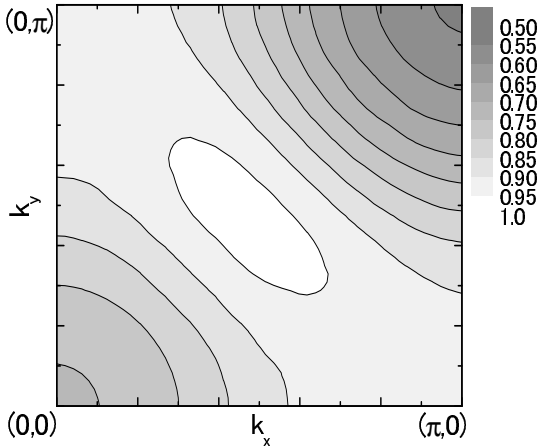}}
\caption{Contour map of the real part of the 
fermion renormalization factor 
Re$Z(\vec k, 0)$ in the off-diagonal coupling with the vertex correction. 
Parameters are chosen to be 
$t/J=3$, $t_1/J=-1.5$, ${\widetilde g}_{off}/J=0.1$, $\omega_{ph}/J=0.5$ 
and $x=0.1$. 
} 
\label{fig:zvtx}
\end{figure}

We present, in Fig.~\ref{fig:zvtx}, a contour map for the renormalization 
factor for fermion Re$Z(\vec k, \omega=0)$ with the vertex correction. 
In the case where the vertex correction is included, 
the essence of the momentum dependence of Re$Z(\vec k, 0)$ pointed 
out in the previous section, 
i.e. the large renormalization around $\vec k=(0,0)$ and $(\pi, \pi)$, 
remains, although Re$Z(\vec k, 0)$ around $\vec k=(\pi,\pi)$ is 
slightly reduced.

\section{Discussion and conclusions}

The interplay between the electron-phonon interaction and 
strong electronic correlation is one of the most important 
issues in many systems such as cuprates, manganites, and C$_{60}$.
Especially in cuprates, it has been often claimed 
that the strong correlation dominates to hide or reduce
the electron-phonon interaction. This seems to be true for the 
diagonal coupling of phonon because the charge density 
fluctuation is suppressed by the Mott physics. 
This is assumed also in the present paper neglecting the 
diagonal electron-phonon coupling. However it also plays an important 
role in the dynamics of the hole observed in the 
ARPES of the undoped cuprates \cite{ronning} because the
photo-doped hole is subject to the rather strong coupling to 
the breathing phonon mode. We believe that this electron-phonon interaction 
is the origin of the anomalously large width of the ARPES spectra. \cite{andrei} 

Now some remarks are in order on other works on the 
electron-phonon interaction in cuprates. 
R$\rm \ddot o$sch and Gunnarsson \cite{rosch}
studied the electron-phonon interaction in the $t-J$ model 
in a similar way estimating the value of 
the coupling constants from the $r$-dependence of 
the transfer integral and the inter-site Coulomb interaction between the 
O and Cu ions ($r$ is the distance between these two
ions). The off-diagonal coupling constant is an order of
magnitude smaller than what we estimated, while the 
diagonal-coupling constant is huge of the order of $0.25$eV.
Therefore they concluded the diagonal coupling is the dominant 
mechanism of electron-phonon interaction in cuprates, even though
it is reduced considerably by the same mechanism we 
discussed in this paper.
However the small value of off-diagonal coupling constant is 
due to the rather subtle cancellation between two terms, and
might change in a more realistic calculation. Also the 
vertex correction strongly enhances the off-diagonal coupling.

Huang {\it et al.} \cite{hanke} recently studied the electron-phonon vertex correction 
for the diagonal coupling due to the Coulomb correlation, 
and found that the large momentum transfer 
scattering is reduced considerably while 
that of small momentum transfer is enhanced. This feature is consistent 
with the earlier analytic studies on the vertex correction. \cite{zeyher}
The important point made in that paper \cite{hanke} is that the interaction 
between the quasi-particle and phonon is {\it enhanced} by the 
electron correlation because of the enhanced density of states. 
This might have an important implication for the 
kink structure observed in the ($\pi$,0) momentum region. \cite{cuk} 

In summary, we have studied the electron-phonon interaction in 
HTSC cuprates motivated by the recent ARPES experiments. 
The off-diagonal electron-phonon coupling, which modulates the hopping 
integral of the Zhang-Rice singlet and the superexchange interaction, 
is relevant in the small doping region. 
We formulate the vertex correction beyond the mean-filed slave boson theory. 
As well as the slave boson and the Langrange multiplier, 
the fluctuations for the fermion hopping and paring lead to the 
collective modes in the superconducting state. \cite{lee2} 
Thus, the dynamics of the collective modes dominates the vertex 
function for the electron-phonon interaction. 
A characteristics in the off-diagonal coupling constant 
$g_{off}^\mu(\vec k, \vec q)$ 
are consistent with several $\vec k$ (electron momentum) and $\vec q$ 
(phonon momentum) dependent experiments; 
the anisotropy in the quasiparticle renormalization 
reflecting the momentum-dependent velocity ratio across the kink and 
its peak-width, and the phonon momentum dominating the strong 
electron-phonon coupling. 
With the vertex correction, the attractive interaction increases 
and the phonon momentum induces the interaction shifts to 
$(\pi(1-\delta),0)$ with $\delta \cong 0.3$. 
These changes are caused by the amplitude fluctuation mode and the 
$\phi$ gauge mode, that is, the collective fluctuations of the order 
parameter amplitude $|\chi_{ij}|^2 +|\Delta_{ij}|^2$, 
and its out-of-phase fluctuation mode, respectively. 
However, the kink position in the ARPES spectra shifts 
from $\omega_{ph}$ to $\omega_{ph}+2 \alpha \Delta$ 
across T$_c$ in our calculation, 
while it does not in the experiments. 
The resolution of this discrepancy is left for future studies. 

It is evident that the strong electron correlation plays the 
major role in the physics of HTSC cuprates, and the role of the
phonon is also modified by it. Some specific channel of the 
electron-phonon interaction is selected to be enhanced by the 
strong correlation, and this will avoid the competition 
with the magnetic mechanism working in the different momentum regions.
Therefore it seems that the electron correlation 
and the electron-phonon interaction seems to collaborate to enhance the
$d$-wave pairing.
The role of the other phonons besides the half-breathing mode 
studied here, is left for future investigations.
\par

\begin{acknowledgments}
We would like to acknouwledge Z.~-X.~Shen, A.~Lanzara, T.~Cuk,  
A.~Fujimori, T.~Egami, M.~Tachiki, W.~Hanke, 
S.~Sugai, and O.~Gunnarsson for their stimulus discussions. 
This work is supported by 
KAKENHI from MEXT. 
Part of the numerical calculation has been performed by 
the supercomputing facilities in IMR, Tohoku University. 
\end{acknowledgments}


\begin{references}
%
\bibitem{baskaran}
G.~Baskaran, Z.~Zou, and P.~W.~Anderson, 
Sol. Stat. Comm. {\bf 63}, 973 (1987). 
\bibitem{kotliar}
G.~Kotliar, and J.~Liu, 
Phys. Rev. Lett. {\bf 61}, 1784 (1988). 
\bibitem{suzumura}
Y.~Suzumura, Y.~Hasegawa, and H.~Fukuyama, 
Jour. Phys. Soc. Jpn. {\bf 57}, 2768 (1988). 
\bibitem{nagaosa}
N.~Nagaosa, and P.~A.~Lee, 
Phys. Rev. Lett. {\bf 64}, 2450 (1990), and 
P.~A.~Lee, and N.~Nagaosa, 
Phys. Rev. B {\bf 46}, 5621 (1992). 
\bibitem{ioffe}
L.~B.~Ioffe, and A.~I.~Larkin, 
Phys. Rev. B {\bf 39}, 8988 (1989). 
%
%
\bibitem{yonemitsu}
K.~Yonemitsu, A.~R.~Bishop, and J.~Lorenzana, 
Phys. Rev. Lett. {\bf 69}, 965 (1992). 
\bibitem{ishihara}
S.~Ishihara, T.~Egami, and M.~Tachiki, 
Phys. Rev. B {\bf 53}, 3163 (1997). 
\bibitem{horsch}
G.~Khaliullin,  and P.~Horsch,  
Phys. Rev. B {\bf 54}, R9600 (1996). 
\bibitem{nazarenko}
A.~Nazarenko, and E.~Dagotto, 
Phys. Rev. B {\bf 53}, R2987 (1996). 
\bibitem{normand}
B.~Normand, H.~Kohno, and H.~Fukuyama, 
Phys. Rev. B {\bf 53}, 856 (1996). 
\bibitem{boeri}
L.~Boeri, E.~Cappelluti, C.~Grimaldi, and L.~Pietroneto, 
Inter. Jour. Mod. Phys. B {\bf 14}, 2970 (2000).
%
\bibitem{crawford}
M.~K.~Crawford, M.~N.~Kunchur, 
W.~E.~Farneth, E.~M.~McCarron III, and S.~J.~Poon, 
Phys. Rev. B {\bf 41}, 282 (1990). 
\bibitem{keller}
G.~-M.~Zhao, M.~B.~Hunt, H.~Keller, and K.~A.~M$\rm \ddot{u}$ller, 
Nature {\bf 385}, 236 (1997), 
and G.-M.~Zhao, H.~Keller, and K.~Conder, 
cond-mat/0204447.  
%
\bibitem{mcqueeney}
R.~J.~McQueeney, Y.~Petrov, T.~Egami, M.~Yethiraj, 
G.~Shirane, and Y.~Endoh, 
Phys. Rev. Lett. {\bf 82}, 628 (1999). 
\bibitem{pintschovius}
L.~Pintschovius, and M.~Braden, 
Phys. Rev. B {\bf 60}, R15~039 (1999), 
and M.~Braden, W.~Reichardt, S.~Shiryaev, and S.~N.~Barilo, 
Physica C {\bf 378-381}, 89 (2002).
\bibitem{yamada}
T.~Fukuda, J.~Mizuki, K.~Ikeuchi, K.~Yamada, M.~Fujita, 
Y.~Endoh, A.~Q.~R.~Baron, T.~Tsutsui, and Y.~Tanaka, 
(unpublished). 
%
\bibitem{bogdanov}
P.~V.~Bogdanov, A.~Lanzara, S.~A.~Keller, X.~J.~Zhou, 
E.~D.~Liu, W.~J.~Zheng, G.~Gu, J.~-I.~Shimoyama, 
K.~Kishio, H.~Ikeda, R.~Yoshizaki, Z.~Hussain, and 
Z.~-X.~Shen, 
Phys. Rev. Lett. {\bf 85}, 2581 (2000). 
\bibitem{kaminski}
A.~Kaminski, 
M.~Randeria, J.~C.~Campuzano, 
M.~R.~Norman, H.~Fretwell, J.~Mesot, 
T.~Sato, T.~Takahashi, and K.~Kadowaki, 
Phys. Rev. Lett. {\bf 86}, 1070 (2001). 
\bibitem{johnson}
P.~D.~Johnson, T.~Valla, A.~V.~Fedorov, Z.~Yusof, B.~O.~Wells, Q.~Li, 
A.~R.~Moodenbaugh, G.~D.~Gu, 
N.~Koshizuka, C.~kendziora, Sha Jian, and D.~G.~Hink, 
Phys. Rev. Lett. {\bf 87}, 177077 (2001). 
\bibitem{lanzara}
A.~Lanzara, P.~V.~Bogdanov, X.~J.~Zhou, S.~A.~Keller, 
D.~L.~Feng, E.~D.~Lu, T.~Yoshida, H.~Eisaki, A.~Fujimori, K.~Kishio, J.~-I.~Shimoyama, 
T.~Noda, S.~Uchida, Z.~Hussain, and Z.~-X.~Shen, 
Nature {\bf 412}, 510 (2001).  
%
\bibitem{shen_pm}
Z.~-X.~Shen, A.~Lanzara, S.~Ishihara, and N.~Nagaosa, 
Phil. Mag. {\bf 82}, 1349 (2002).  
%
\bibitem{tachiki}
M.~Tachiki, M.~Machida, and T.~Egami, 
Phys. Rev. B {\bf 67}, 174506 (2003). 
\bibitem{hanke}
Z.~B.~Huang, W.~Hanke, E.~Arigoni, and D.~J.~Scalapino, 
cond-mat/03006131. 
\bibitem{carbotte}
E.~Schachinger, J.~J.~Tu, and J.~P.~Carbotte, 
Phys. Rev. B {\bf 67}, 214508 (2003). 
\bibitem{rosch}
O.~R$\rm \ddot o$sch, and O.~Gunnarsson, 
cond-mat/0308035. 
%
\bibitem{basov}
A.~V.~Puchekov, P.~Fournier, D.~N.~Basov, T.~Timusk, 
A.~Kapitulnik, and N.~N.~Kolesnikov,  
Phys. Rev. Lett. {\bf 77}, 3212 (1996), 
and D.~N.~Basov, R.~Liang, B.~Dabrowski, D.~A.~Bonn, W.~N.~Hardy, and T.~Timusk, 
Phys. Rev. Lett. {\bf 77}, 4090 (1996). 
\bibitem{singley}
E.~J.~Singley, D.~N.~Basov, K.~Kurahashi, T.~Uefuji and K.~Yamada, 
Phys. Rev. B {\bf 64}, 224503 (2003). 
%
\bibitem{zhou}
X.~J.~Zhou, T.~Yoshida, A.~Lanzara, P.~V.~Bogdanov, 
S.~A.~Keller, K.~M.~Shen, W.~L.~Yang, F.~Ronning, T.~Sasagawa, 
T.~Kakeshita, T.~Noda, H.~Eisaki, S.~Uchida, C.~T.~Lin, 
F.~Zhou, J.~W.~Xiang, W.~X.~Ti, Z.~X.~Zhao, A.~Fujimori, Z.~Hussain, and Z.~-X.~Shen, 
Nature {\bf 423}, 398 (2003). 
%
\bibitem{eschrig}
M.~Eschrig, and M.~R.~Norman, 
Phys. Rev. Lett. {\bf 85}, 3261 (2000). 
\bibitem{manske}
D.~Manske, I.~Eremin, and K.~H.~Bennemann, 
Phys. Rev. Lett. {\bf 87}, 177055 (2003). 
%
\bibitem{lanzara_iso}
A.~Lanzara, {\it et al.} (unpublished). 
%
\bibitem{gromko}
A.~D.~Gromko, A.~V.~Fedrov, Y.~-D.~Chuang, J.~D.~Koralek, 
Y.~Aiura, Y.~Yamaguchi, K.~Oka, Y.~Ando, and D.~S.~Dessau, 
cond-mat/0202329.
\bibitem{takahashi}
T.~Satoh, T.~Takahashi, H.~Ding, H.~-B.~Yang, S.~-C.~Wang, T.~Fujii, 
T.~Watanabe, A.~Matsuda, T.~Terashima, and K.~Kadowaki, 
Phys. Rev. Lett. {\bf 91}, 157003 (2002). 
\bibitem{cuk}
T.~Cuk, F.~Baumberger, D.~H.~Lu, N.~Ingle, 
X.~J.~Zhou, H.~Eisaki, N.~Kaneko, Z.~Hussain, N.~Nagaosa, 
and Z.~-X.~Shen, 
(unpublished). 
%
\bibitem{affleck}
I.~Affleck, Z.~Zou, T.~Hsu, and P.~W.~Anderson, 
Phys. Rev. B {\bf 38}, 745 (1988). 
\bibitem{wen}
X.~-G.~Wen, and P.~A.~Lee, 
Phys. Rev. Lett. {\bf 76}, 503 (1996). 
\bibitem{lee}
P.~A.~Lee, N.~Nagaosa, T.~-K.~Ng, and 
X.~-G.~Wen, 
Phys. Rev. B {\bf 57}, 6003 (1998). 
\bibitem{brinckmann}
J.~Brinckmann, and P.~A.~Lee, 
Phys. Rev. B {\bf 65}, 014502 (2001). 
\bibitem{lee2}
P.~A.~Lee, and N.~Nagaosa, 
Phys. Rev. B {\bf 68}, 024516 (2003). 
%
\bibitem{zeyher}
R.~Zeyher, and M.~L.~Kulic, 
Phys. Rev. B {\bf 53}, 2850 (1996). 
\bibitem{kim}
J.~H.~Kim, K.~Levin, R.~Wentzcovitch, and A.~Auervach, 
Phys. Rev. B {\bf 44}, 5148 (1991). 
\bibitem{zhang}
F.~C.~Zhang, and T.~M.~Rice, 
Phys. Rev. B {\bf 37}, 3759 (1988). 
\bibitem{lorenzana}
J.~Lorenzana, and G.~A.~Sawatzky, 
Phys. Rev. Lett. {\bf 74}, 1876 (1995). 
%
\bibitem{norman}
M.~R.~Norman, M.~Randeria, H.~Ding, and J.~C.~Campuzano, 
Phys. Rev. B {\bf 52}, 615 (1995).
%
\bibitem{dewilde}
Y.~DeWilde, N.~Miyakawa, P.~Guptasarma, M.~Iavarone, 
L.~Ozyuzer, J.~F.~Zasadzinski, P.~Romano, 
D.~G.~Hinks, C.~Kendziora, G.~W.~Crabtree, and K.~E.~Gray, 
Phys. Rev. Lett. {\bf 80}, 153 (1998). 
\bibitem{renner}
Ch.~Renner, B.~Revaz, J.~-Y.~Genoud, K.~Kadowaki, and $\rm \O$.~Fischer, 
Phys. Rev. Lett. {\bf 80}, 149 (1998). 
%
\bibitem{gotze}
W.~G$\rm \ddot o$tze, and P.~W$\rm \ddot o$lfle, 
Phys. Rev. B {\bf 6}, 1226 (1972). 
\bibitem{takagi}
H.~Takagi, B.~Batlogg, H.~L.~Kao, J.~Kow, R.~J.~Cava, 
J.~J.~Krajewski, and W.~F.~Peck,~Jr, 
Phys. Rev. Lett. {\bf 69}, 2975 (1992). 
\bibitem{bogdanov2}
P.~V.~Bogdanov, A.~Lanzana, X.~J.~Zhou, W.~L.~Yang, H.~Eisaki, Z.~Hussain, 
and Z.~-X.~Shen, 
Phys. Rev. Lett. {\bf 89}, 167002 (2002). 
%
\bibitem{marston}
J.~B.~Marston, and I.~Affleck, 
Phys. Rev. B {\bf 39}, 11~538 (1989). 
\bibitem{read}
N.~Read, and E.~Sachdev, 
Phys. Rev. Lett. {\bf 62}, 1694 (1989). 
%
\bibitem{Loram}
J.~W.~Loram, K.~A.~Mirza, J.~M.~Wade, J.~R.~Cooper, and W.~Y.~Liang, 
Physica C {\bf 235-240}, 134 (1994), and 
J.~W.~Loram, K.~A.~Mirza, J.~R.~Cooper, and J.~L.~Tallon, 
Physica C {\bf 282-287}, 1405 (1997). 
\bibitem{ronning}
F.~Ronning, C.~Kim, D.~L.~Feng, D.~S.~Marshall, 
A.~G.~Loeser, L.~L.~Miller, J.~N.~Eckstein, 
I.~Bozovic, and Z.~-X.~Shen,  
Science {\bf 282}, 2067 (1998). 
\bibitem{andrei} 
A.~S.~Mishchenko {\it et al.}, (unpublished).
%
\end{references}
\end{document}